\documentclass[12pt]{iopart}

\usepackage{graphicx}
\begin{document}

\title[Author guidelines for IOP journals in  \LaTeXe]{Turbulence in quantum fluids}

\author{Makoto Tsubota}

\address{
Department of Physics, Osaka City University, Sumiyoshi-ku, Osaka, 558-8585,  Japan}
\ead{tsubota@sci.osaka-cu.ac.jp}
\begin{abstract}
This paper reviews briefly the recent important developments in the physics of quantum turbulence (QT) in superfluid helium and atomic Bose-Einstain condensates (BECs). 
After giving basics of quantum hydrodynamics, we discuss energy spectrum, QT created by vibrating structures, visualization among topics on superfluid helium.
For atomic BECs we review three-dimensional QT, two-component BECs, and spin turbulence in spinor BECs.
The last part is devoted to some perspectives of this issue.

\end{abstract}

\maketitle

\section{Introduction}
Bose-Einstein condensation is often considered to be a macroscopic quantum phenomenon. 
This is because bosons constituting a system occupy the same single-particle ground state below a critical temperature through Bose-Einstein condensation to form a macroscopic wave function (order parameter) extending over the entire system.
As the direct result, quantized vortices appear in the Bose-condensed system.  
A quantized vortex is a vortex of inviscid superflow, and any rotational motion of a superfluid is sustained by quantized vortices.
Hydrodynamics dominated by quantized vortices is called {\it quantum hydrodynamics} (QHD), and turbulence comprised of quantized vortices is known as {\it quantum turbulence} (QT). 
The studies of quantized vortices originally began in 1950s using superfluid $^4$He, and much theoretical, numerical, and experimental effort has been devoted to the field.

Research into QHD and QT has tended toward new directions since the mid 1990s \cite{PLTP,TsubotaJPSJ}.  
One new direction is in the field of low temperature physics, studying superfluid helium. 
This field of study started with the attempt to understand the relationship between QT and {\it classical turbulence} (CT) \cite{Vinen2010}.  
The energy spectrum of fully developed CT is known to obey the Kolmogorov law in the inertial range \cite{Frisch,Davidson}.  
Recent experimental and numerical studies support a Kolmogorov spectrum in QT. 
Following these studies, the research of QT on superfluid helium has progressed to important topics such as the dissipation process at very low temperatures, QT created by vibrating structures, and the visualization of QT \cite{PLTP}.
Another new direction is the realization of Bose--Einstein condensation in trapped atomic gases, first performed in 1995, which has stimulated intense experimental and theoretical activity \cite{Pethick2008, Pitaevskii2003}.  
As proof of the existence of superfluidity, quantized vortices have been created and observed in atomic Bose-Einstein condensates (BECs), and considerable effort has been devoted to a number of fascinating problems related to this.  
Atomic BECs have several advantages over superfluid helium.  
The most important is that modern optical techniques enable the direct control and the visualization of  condensates. 

The present article will review these recent studies on QHD and QT in superfluid helium and atomic BECs.
Since the topics are diverse, we do not intend any all-inclusive article.
The readers should refer to other recent review articles for the overview of this field \cite{PLTP,TsubotaJPSJ,Vinen2010, Skrbekreview, Novelsuperfluids, PR}. 
The aim of this article is to address the essential ideas of this issue and the strength of this topic.  
Section 2 describes the minimum information of this field necessary for understanding the following parts.
Section 3 is the main part, reviewing the characteristics of this system with referring to some recent important results.  
Section 4 discusses the perspectives of studying quantum hydrodynamics and turbulence.

\section{Basics of quantum hydrodynamics}
This section descricbes the basics on BEC and superfluidity necessary for understanding Sec. 3.

\subsection{Bose-Einstein condensation, and quantized vortices}
Quantum hydrodynamics has been studied in connection with superfluid helium for over 50 years.
Liquid $^4$He enters a superfluid state below the $\lambda$ point (2.17 K) through
Bose--Einstein condensation of the atoms. 
The characteristics of superfluidity were discovered experimentally
in the 1930s by Kapitza {\em et al.}
The hydrodynamics of superfluid helium is well described by the two-fluid model,
in which the system consists of an inviscid superfluid (of density $\rho_s$) and 
a viscous normal fluid (of density $\rho_n$) with two independent velocity fields $\mathbf{v}_s$ and $\mathbf{v}_n$. 
The mixing ratio of the two fluids depends on temperature. 
As the system is cooled down below the $\lambda$ point, the ratio of 
the superfluid component increases, and the entire fluid becomes
superfluid below about 1 K.
The Bose-condensed system exhibits the macroscopic wavefunction
$\Psi(\mathbf{x},t)=|\Psi(\mathbf{x},t)| e^{i \theta(\mathbf{x},t)}$ as an order parameter. 
The superfluid velocity field representing the potential flow is given by $\mathbf{v}_s=(\hbar/m) \nabla \theta$ with boson mass $m$. 
Since the macroscopic wavefunction must be single-valued for space coordinate $\mathbf{x}$, the circulation $\Gamma = \oint \mathbf{v} \cdot d\mathbf{\ell}$ for an arbitrary closed loop within the fluid is quantized in terms of the quantum value $\kappa=h/m$.  A vortex with such quantized circulation is called a quantized vortex. Any rotational motion of a superfluid is only sustained 
by quantized vortices.

A quantized vortex is a topological defect characteristic of a Bose--Einstein condensate and is different from a vortex in a classical viscous fluid. 
First, its circulation is quantized and conserved, in contrast to a classical vortex whose circulation can have any value and is not conserved. 
Second, a quantized vortex is a vortex of inviscid superflow. Thus, it cannot decay by the viscous diffusion of vorticity that occurs in a classical fluid.  
Third, the core of a quantized vortex is very thin, on the order of the coherence length, which is only a few angstroms in superfluid $^4$He and submicron in atomic BECs.
Since the vortex core is thin and does not decay by diffusion, it is possible to identify the position of a quantized vortex in the fluid. 
These properties make a quantized vortex more stable and definite than a classical vortex. 

\subsection{Brief research history of superfluid helium}
Early experimental studies of superfluid turbulence focused primarily on thermal counterflow, in which the normal fluid and superfluid flow in opposite directions. 
The flow is driven by an injected heat current, and it is found that the superflow becomes dissipative when the relative velocity between the two fluids exceeds a critical value \cite{GorterMellink}. 
Feynman proposed that it is a superfluid turbulent state consisting of a tangle of quantized vortices \cite{Feynman}. 
Vinen later confirmed Feynman's proposal experimentally by showing that the dissipation arises from mutual friction between vortices and the normal flow \cite{Vinen57a, Vinen57b, Vinen57c, Vinen58}. 
The quantization of circulation was also observed by Vinen using a vibrating wire technique \cite{Vinen61}. 
Subsequently, many experimental studies have examined superfluid turbulence (ST) in thermal counterflow systems, revealing a variety of physical phenomena \cite{Tough82}. 
The dynamics of quantized vortices are nonlinear and nonlocal, so it has not been easy to quantitatively understand the experimental results on the basis of vortex dynamics. 
A breakthrough was achieved by Schwarz, who clarified the picture of ST consisting of tangled vortices by a numerical simulation of the quantized vortex filament model in the thermal counterflow \cite{Schwarz85, Schwarz88}. 
However, since the thermal counterflow has no analogy in conventional fluid dynamics, these experimental and numerical studies are not helpful in clarifying the relationship between ST and classical turbulence (CT). 
Superfluid turbulence is often called quantum turbulence (QT), which emphasizes the fact that it is comprised of quantized vortices.

\subsection{Model of quantum turbulence}
There are generally two models available for the numerical simulation of quantum turbulence.
One is the vortex filament model and the other is the Gross-Pitaevskii (GP) model.  

\subsubsection{Vortex filament (VF) model}
As discussed in Sec. 2.1, a quantized vortex has quantized circulation. 
The vortex core is extremely thin, usually much smaller than other characteristic length scales of the vortex motion. 
These properties allow a quantized vortex to be represented as a vortex filament. 
In classical fluid dynamics, the vortex filament (VF) model is only an idealization. 
However, the VF model is accurate and realistic for a quantized vortex in superfluid helium.

The VF model represents a quantized vortex as a filament passing through the fluid, having a definite direction corresponding to its vorticity. 
Except for the thin core region, the superflow velocity field has a classically well-defined meaning and can be described by ideal fluid dynamics \cite{Schwarz85}. 
The velocity at point $\mathbf{r}$ due to a filament is given by the Biot--Savart expression,
\begin{equation}
\mathbf{v}_{s} (\mathbf{r} )=\frac{\kappa}{4\pi}\int_{\cal L} \frac{(\mathbf{s}_1 - \mathbf{r}) \times
d\mathbf{s}_1}{|\mathbf{s}_1-\mathbf{r}|^3}.
\label{BS}
\end{equation}
The vortex moves with the superfluid velocity. At finite temperatures, mutual friction occurs between the vortex core and the normal flow. 

A numerical simulation method on this model has been described in detail elsewhere \cite{Schwarz85, Schwarz88,Tsubota00}. 
A vortex filament is represented by a single string of points separated by distance $\Delta\xi$. 
The vortex configuration at a given time determines the velocity field in the fluid, thus causing the vortex filaments to move. 
Vortex reconnection needs to be included when simulating vortex dynamics.  
A numerical study of a classical fluid shows that close interaction of two vortices leads to their reconnection, chiefly because of the viscous diffusion of vorticity. 
Schwarz assumed that two vortex filaments reconnect when they come within a critical distance of each other and showed that statistical quantities such as the vortex line density are not sensitive to how the reconnections occur \cite{Schwarz85, Schwarz88}.  
Even after Schwarz's study, it remained unclear as to whether quantized vortices can actually reconnect. 
However, Koplik and Levine directly solved the GP model to show that two closely quantized vortices reconnect even in an inviscid superfluid \cite{Koplik93}. 
Later simulations have shown that reconnections are accompanied by the emission of sound waves having wavelengths on the order of the healing length \cite{Leadbeater01, Ogawa02}. 

\subsubsection{The Gross-Pitaevskii (GP) model}
In a weakly interacting Bose system, the macroscopic wavefunction $\Psi(\mathbf{x},t)$ appears as the order parameter of the Bose--Einstein condensation, obeying the GP equation \cite{Pethick2008},
\begin{equation}
i \hbar \frac{\partial \,\Psi(\mathbf{x},t)}{\partial \,t} = \biggl( - \frac{\hbar ^2}{2m}\nabla^2
+g |\Psi(\mathbf{x},t)|^{2}-\mu \biggr) \Psi(\mathbf{x},t).
\label{gpeq}
\end{equation}
Here $g$ refers to the interaction parameter and $\mu$ is the chemical potential.
Writing $\Psi = | \Psi | \exp (i \theta)$, the squared amplitude $|\Psi|^2$ is the condensate density and the gradient of the phase $\theta$ gives the superfluid velocity  $\mathbf{v}_s = (\hbar/m) \nabla \theta$, corresponding to frictionless flow of the condensate. 
The only characteristic scale of the GP model is the coherence length defined by $\xi=\hbar/(\sqrt{2mg}\,| \Psi |)$, which determines the vortex core size. 

\subsubsection{VF model or GP model}
This subsection describes briefly which model should be used in realistic simulation. 
The criterion comes from advantages and disadvantages of each model. 
The VF model neglects the core structure of quantized vortices. 
Hence, it can describe well the dynamics of quantized vortices as far as the coherence length is much smaller
 than other characteristic lengths such as the typical radius of curvature of vortex lines, the mean distance $\ell$ between vortices, and the system size.
However, it cannot explain phenomena related to vortex cores, such as reconnection, nucleation and annihilation.
Thus the VF model is usually used for the simulation in superfluid $^4$He and superfluid $^3$He-B.
On the other hand, the GP model can describe not only the dynamics of vortices but also the phenomena concerned with vortex cores, 
being useful for trapped atomic BECs in which the coherence length is usually not much smaller than the system size.
The GP model is not quantitatively applicable to superfluid helium which is a strongly correlated system.
However, the GP model has long been studied as the simplest model of BEC and superfluidity, actually leading to lots of fruitful results.

\section{Recent important developments on the studies of quantum turbulence}
The study of quantized vortices and QT has increased in intensity for two reasons.  
The first is that recent studies of QT are considerably advanced over older studies, which were chiefly limited to thermal counterflow in $^4$He, which has no analogue with classical traditional turbulence,  
whereas new studies on QT are focused on a comparison between QT and CT.
The second reason is the realization of atomic BECs in 1995, 
for which modern optical techniques enable the direct control and visualization of the condensate and can even change the interaction; such direct control is impossible in other quantum condensates like superfluid helium and superconductors. 
This section reviews briefly the recent research developments in these two fields.

\subsection{Recent important results in superfluid helium}
Most older experimental studies on QT were devoted to thermal counterflow. 
Since this flow has no classical analog, these studies did not significantly contribute to the understanding of the relation between CT and QT. 
Since the mid 1990s, important experimental studies have been published on QT in the absence of thermal counterflow, differing significantly from previous studies \cite{TsubotaJPSJ,Vinen2010}.
One of the important interests is whether the energy spectrum $E(k)$ of QT exhibits the Kolmogorov law 
\begin{equation}
E(k)=C\epsilon ^{2/3}k^{-5/3},
\label{K41}\end{equation}
where $\epsilon$ is the energy flux and $k$ is the wave number. 
The Kolmogorov law is the most important statistical law in turbulence \cite{Frisch,Davidson}.

The first important contribution was made by Maurer and Tabeling \cite{Maurer98}, who confirmed experimentally the Kolmogorov spectrum in superfluid $^4$He.
A turbulent flow was produced in a cylinder by driving two counter-rotating disks.
The authors measured the local pressure fluctuations to obtain the energy spectrum.  
The experiments were made at three different temperatures: 2.3 K, 2.08 K, and 1.4 K.
In each case, the Kolmogorov spectrum (\ref{K41}) was confirmed.

Next was a series of experiments on grid turbulence performed for superfluid $^4$He above 1 K by the Oregon 
group \cite{Stalp99, Skrbek00a,Skrbek00b,Stalp02}.  
Flow through a grid is usually used for generating turbulence in classical fluid dynamics.
At sufficient distance behind the grid, the flow displays homogeneous isotropic turbulence.
This method has also been applied to superfluid helium.  
In the Oregon experiments, the helium was contained in a channel
along which a grid was pulled at constant velocity. 
A pair of second-sound transducers was implanted in the walls of the channel to observe vortex tangles.
In combining their observations with the decay of the turbulence, the authors had to make several assumptions.
The analysis is too complicated to be described in detail here.  
The important point is that the coupling between the superfluid and the normal fluid by mutual friction makes a quasiclassical flow appear 
at length scales much larger than the mean intervortex spacing $\ell$ and causes the fluid to behave like a one-component fluid \cite{Vinen00}. 
The vortex line length density $L$ (the vortex line length per unit volume) is found to decay as $t^{-3/2}$.  
A simple analysis shows that the  $t^{-3/2}$ decay can be derived from a quasiclassical model with a Kolmogorov spectrum \cite{Stalp99}.

\subsubsection{Energy spectrum of quantum turbulence}

After these pioneering experimental works, lots of theoretical and numerical efforts reveal the energy spectrum of QT.
Since QT at finite temperatures is much affected by normal fluid and complicated,  we will confine ourselves here to QT at zero temperature. 

\begin{figure}[htb]
\centering
\includegraphics[width=0.75\linewidth]{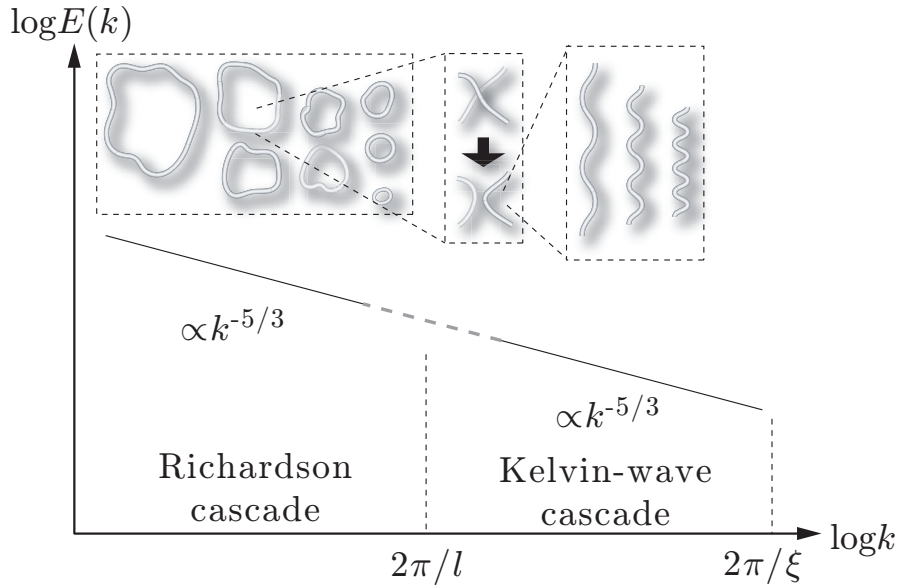}
\caption{\label{fig-two-cascade} Overall picture of the energy spectrum of QT at zero temperature.
The energy spectrum depends on the scale, and its properties change at about the scale of the mean intervortex spacing $l$.
When $k < k_l = 2 \pi / l$, a Richardson cascade of quantized vortices transfers energy from large to small scales, maintaining the Kolmogorov spectrum $E(k) = C \varepsilon^{2/3} k^{-5/3}$.
When $k > k_l$, energy is transferred by the Kelvin-wave cascade, which is a nonlinear interaction between Kelvin waves of different wavenumbers.
Eventually, energy is dissipated at scales of $\xi$ by the radiation of phonons.}
\end{figure}
We summarize the overall picture of the energy spectrum of QT at zero temperature on the basis of theoretical and numerical studies (Fig. \ref{fig-two-cascade}) \cite{PLTP}.
In QT, vortices form a complicated tangled structure.
If the tangle is assumed to be homogeneous and isotropic, there are two characteristic length scales: one is the mean intervortex spacing $l = L^{-1/2}$ with a vortex line length density $L$, and the other is the coherence length $\xi$ corresponding to the size of the vortex core.
Generally, $l$ is much larger than $\xi$; $l \gg \xi$.
Using the length scales $l$ and $\xi$, we can define the characteristic wave numbers $k_l = 2 \pi / l$ and $k_{\xi} = 2\pi / \xi$.

At length scales larger than $l$, the dynamics of QT are dominated by a tangled structure of many vortices.
Because vortex dynamics becomes collective at large scales, quantization of the circulation is not relevant, and the dynamics are similar to those of eddies in CT.
As a result, the energy spectrum $E(k)$ in the range of $k < k_l$ obeys the Kolmogorov law  (\ref{K41}).
Vortices in QT sustain a Richardson cascade that transfers energy from smaller wavenumbers to larger ones without dissipation.
The Richardson cascade can be understood as large vortices breaking up into smaller ones in real space.
In the Richardson cascade process, the key vortex dynamics is reconnection; when two vortex lines approach each other, they become locally antiparallel and reconnect \cite{Koplik93,Leadbeater01,Ogawa02}.
Reconnection causes topological changes of the vortex lines in the tangle, formations of distortion waves on the vortex lines (Kelvin waves), or fission of vortex loops through self-reconnections, all of which are responsible for the cascade of energy towards smaller scales \cite{Tsubota00,Svistunov1995,Kivotides2001a}.
The Richardson cascade and the Kolmogorov energy spectrum have been confirmed by several numerical studies of the VF mode and the GP model \cite{Nore1997a,Nore1997b,Araki2002,Kobayashi2005a,Kobayashi2005b,Yepez2009}.
This wavenumber region may be called semi-classical region, because QT mimics CT here.

At length scales smaller than $l$, which is referred to as the quantum region, the Richardson cascade is no longer dominant, and the quantized circulation and the motion of each vortex line becomes significant \cite{Svistunov1995,Mitani2003,Kozik2004,Kozik2005,L'vov10,Laurie10}.
In this range, vortex dynamics are characterized by the cascade process of the Kelvin-waves which are formed by reconnection.
The nonlinear interaction of the Kelvin waves is the origin of the cascade from smaller to larger  wavenumbers.
The cascade dynamics of the Kelvin waves along a single vortex line has been confirmed, both numerically and theoretically.
The Kelvin-wave cascade in QT has also been studied.
The energy spectrum in the quantum region $k_l < k < k_{\xi}$ is theoretically predicted to obey a Kolmogorov-like power law: $E(k) \propto k^\eta$.
There have been serious arguments for the exponent $\eta$.
Kozik and Svistunov (KS) found the spectrum $E_{\rm{KS}}(k) \propto \epsilon^{1/5}k^{-7/5}$ \cite{Kozik2004} with the $k$-independent energy flux, while another spectrum $E_{\rm{LN}}(k) \propto \epsilon^{1/3}k^{-5/3}$ was obtained by L'vov and Nazarenko (LN) \cite{L'vov10}.    
Both KS and LN approaches  come from weak wave turbulence \cite{Nazarenko} of Kelvin waves.
Assuming that Kelvin waves are excited along a straight vortex, the Hamiltonian of wave interaction is represented in series of small wave amplitudes. 
Due to the one-dimensional nature of Kelvin waves the conservation laws for energy and momentum forbid energy exchange in the 2 $\leftrightarrow$ 2 wave scattering.
Only higher-order processes starting from the 3 $\leftrightarrow$ 3 wave scattering cause the energy exchange between Kelvin waves and thus relevant for the energy cascade.
KS showed that the 3 $\leftrightarrow$ 3 interaction amplitude $\cal W$ is proportional to $k^2$ with the Kelvin wave number $k$, which leads to the spectrum $E_{\rm{KS}}(k)$.
This interaction amplitude supposes that the energy cascade is dominated by local, step-by step energy transfer by interacting Kelvin waves of $k$ with those of same order of magnitude $k$. 
However, LN showed the energy cascade was actually nonlocal with ${\cal W} \propto k$, which justified the spectrum $E_{\rm{LN}}(k)$  rather than  $E_{\rm{KS}}(k)$ \cite{Lebedev10}.

In the region $k \sim k_\xi$, the Kelvin waves with wavelength of $\xi$ change to elementary excitations, such as phonons and rotons, in the primary decay process of QT near zero temperature \cite{Vinen2001}.

\subsubsection{QT created by vibrating structures}
Recently, vibrating structures, such as discs, spheres, grids, and wires, have been widely used for research into QT \cite{SkrbekPLTP}. 
Despite detailed differences between the structures considered, the experiments show some surprisingly common phenomena. 

This trend started with the pioneering observation of QT on an oscillating microsphere by J\"ager {\it et al.}\cite{Schoepe1995}. 
The sphere used by J\"ager {\it et al.} had a radius of approximately 100 $\mu$m, 
and was made from a strongly ferromagnetic material with a very rough surface. 
The sphere was magnetically levitated in superfluid $^4$He and its response with respect to the alternating drive was observed.  
At low drives, the velocity response $v$ was proportional to the drive $F_D$, taking the "laminar" form $F_D=\lambda(T) v$, with the temperature-dependent coefficient $\lambda(T)$.  
At high drives, the response changed to the "turbulent" form $F_D=\gamma(T) (v^2-v_0^2)$ above the critical velocity $v_0$. 
At relatively low temperatures the transition from laminar to turbulent response was accompanied by significant hysteresis.
Subsequently, several groups have experimentally investigated the transition to turbulence in the superfluids $^4$He and $^3$He-B by using grids, wires  and tuning forks.  
The details of the observations are described in a review article \cite{SkrbekPLTP}. 
Here we shall briefly describe a few important points necessary for the current article.

These experimental studies reported some common behavior independent of the details of the structures, such as the type, shape, and surface roughness. 
The observed critical velocities are in the range from 1 mm/s to approximately 200 mm/s.  
Since the velocity is usually much lower than the Landau critical velocity of approximately 50 m/s, the transition to turbulence should come not from intrinsic nucleation of vortices but from the extension or amplification of remnant vortices. 
Such behavior is shown in the numerical simulation by the vortex filament model \cite{Hanninen2007, Goto2008, Fujiyama2009}.  

\begin{figure}
	\begin{center}
		\begin{tabular}{cc}		
			\includegraphics[width=0.4\linewidth]{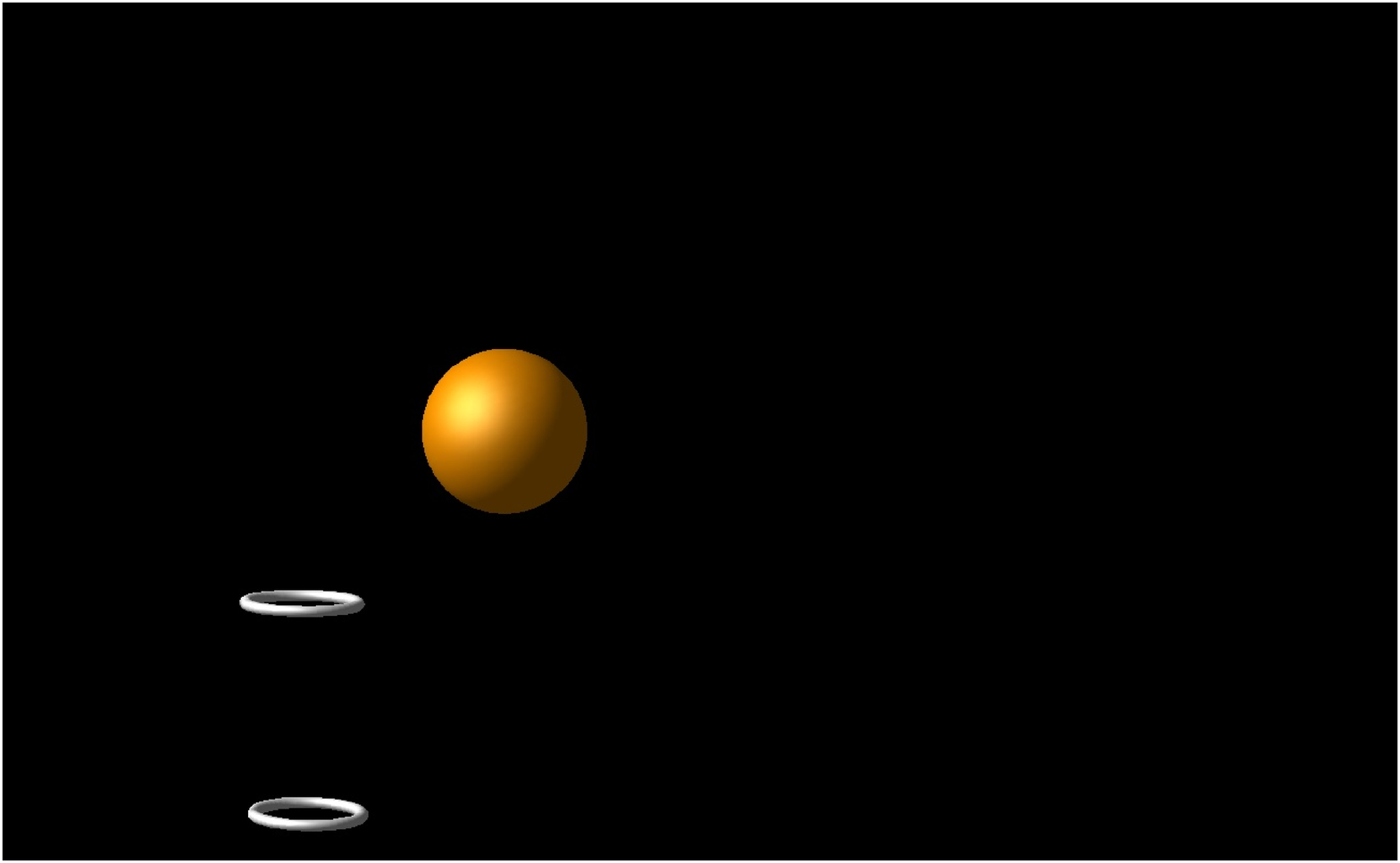}	&
			\includegraphics[width=0.4\linewidth]{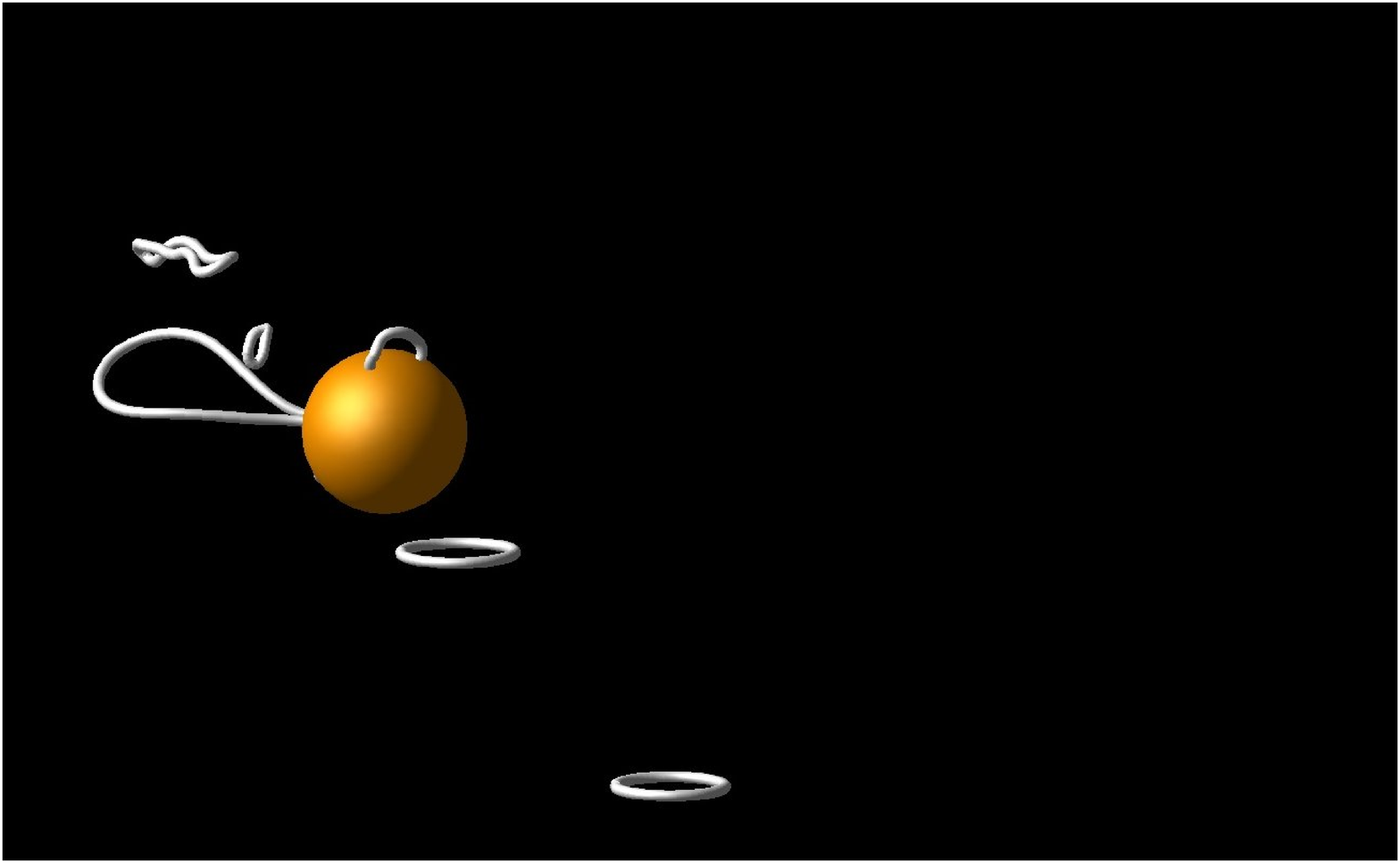}	\\
			(a) $t=19$ ms	&	(b) $t=40$ ms	\\
			\includegraphics[width=0.4\linewidth]{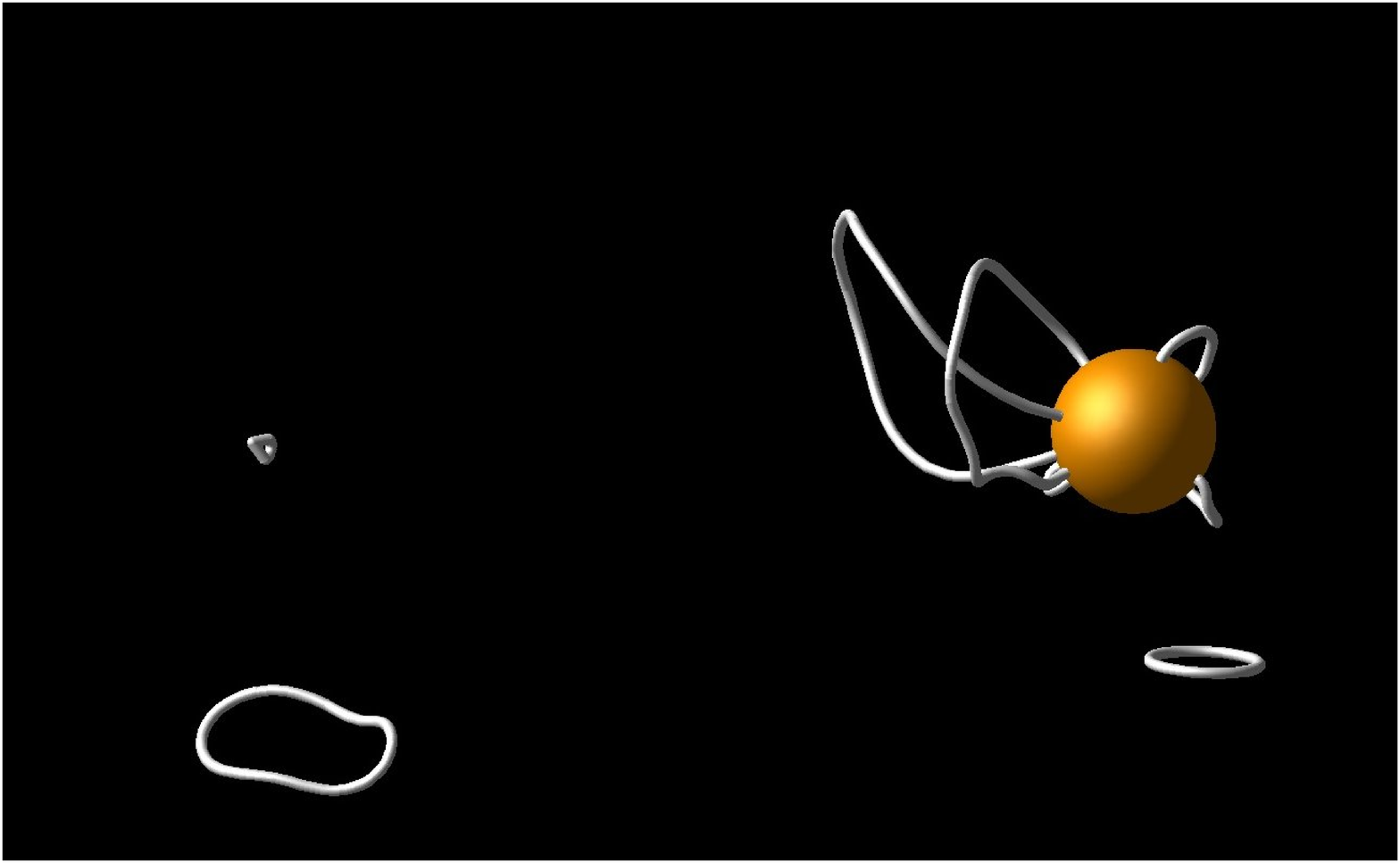}	&
			\includegraphics[width=0.4\linewidth]{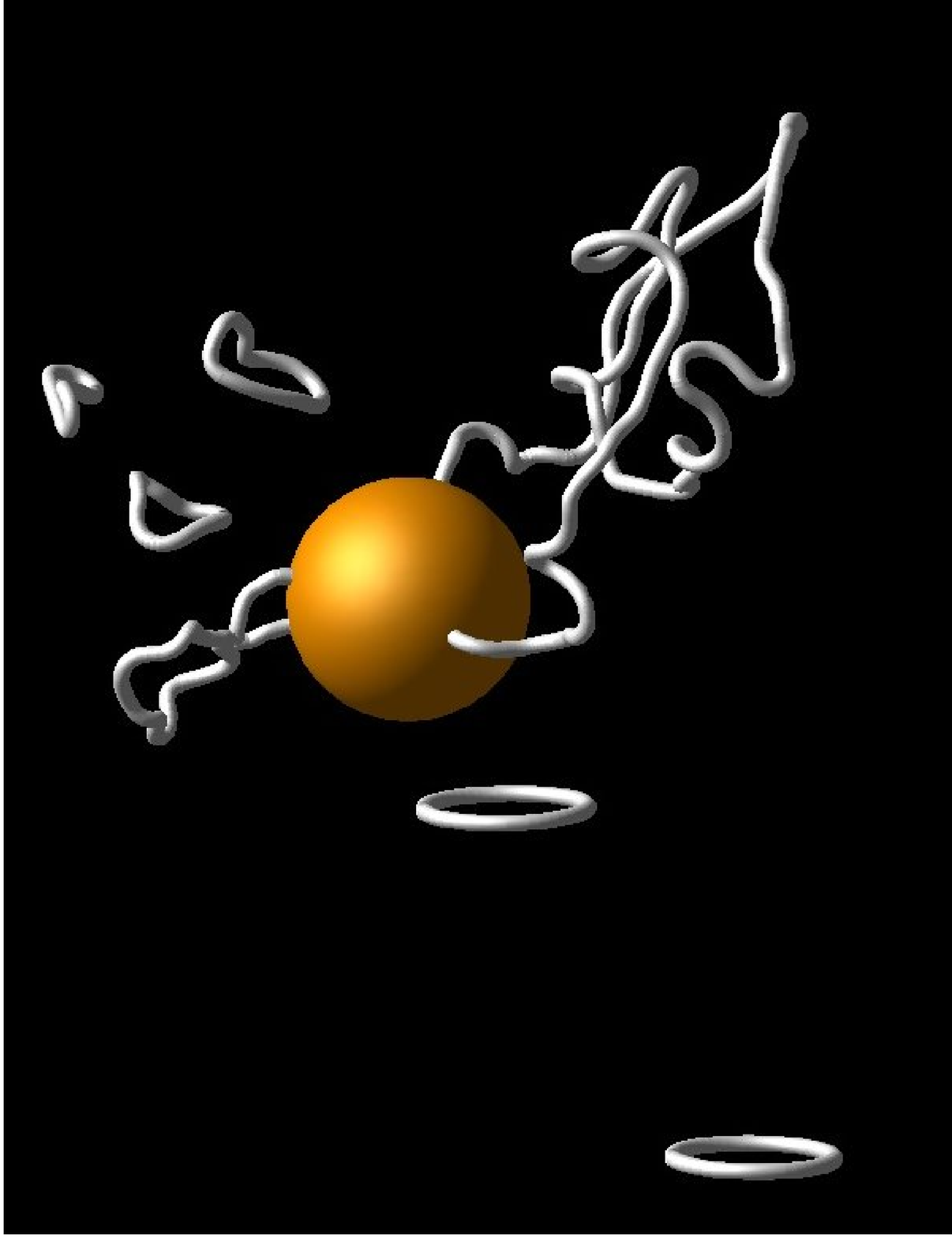}	\\
			(c) $t=58$ ms	&	(d) $t=100$ ms
		\end{tabular}
		\caption{Time evolution of turbulence generation for the case of a sphere oscillating with a velocity magnitude of 90 mm/s. See the text for details. [Fujiyama and Tsubota: Phys. Rev. B \textbf{79} (2009) 094513, reproduced with permission. Copyright 2009, the American Physical Society.]}
		\label{Fujiyama}
	\end{center}
\end{figure}

Generally it is not easy to control the remnant vortices in an actual experimental setup.
However, Goto {\it et al.} succeeded in preparing a vibrating wire free from remnant vortices \cite{Goto2008}.
This wire never causes a transition to turbulence by itself; it can cause turbulence only when it receives seed vortices from another wire.
Such a simulation was performed by Fujiyama {\it et al.} as shown in Fig. \ref{Fujiyama}  \cite{Fujiyama2009}.
The sphere oscillates horizontally; the diameter of the sphere is 3 $\mu$m, the frequency of the oscillation is 1590 Hz, while the oscillation velocity is chosen in the range of 30--90 mm/s.
Vortex rings of radius 1 $\mu$m are injected from the bottom of the medium [Fig. \ref{Fujiyama}(a)].
When the vortex rings collide with the sphere, reconnections occur and the vortices become attached to the sphere.
Then, the attached vortices are stretched as the sphere moves [Fig. \ref{Fujiyama}(b) and (c)].
Due to the successive injection of vortex rings the process is repeated and the stretched vortices form a tangle around the sphere  [Fig. \ref{Fujiyama}(d)].
The vortices grow in size and some then detach from the sphere.
In spite of the detachment, the oscillating sphere still sustains the vortex tangle when its velocity is relatively large. 
The vortex line length in a finite volume including the sphere was calculated at different oscillation velocities as a function of elapsed time.
The loss of the vortices escaping from the volume balances the injection and the growth of the vortices so that the line length saturates.
Only a slight increase in the line length can be observed  for a velocity magnitude of 30 mm/s, which means that the vortices are not stretched by the sphere.
The saturated value of the line length increases with the oscillation velocity magnitude.
For velocity magnitudes above 50 mm/s, the saturated line length value is much larger than the injection of vortices, which suggests that vortex tangles are formed around the sphere.
This behavior is qualitatively consistent with the observations \cite{Goto2008}.

In order to characterize the transition to turbulence, Fujiyama {\it et al.} also studied the drag force \cite{Fujiyama2009}.
The drag force acting on an object in a uniform flow is generally represented by
\begin{equation} F_D=\frac12 C_D \rho A U^2, \label{Drag} \end{equation}
where $C_D$ is the drag coefficient, $\rho$ is the fluid density, $A$ is the projection area of the object normal to the flow, and $U$ is the flow velocity.
It is known in classical fluid mechanics how $C_D$ depends on the properties of the flow.
At low Reynolds number, Stokes's drag force acts on the object, which is proportional to the magnitude of $U$, with the result that $C_D$ is inversely proportional to $U$.
When the flow becomes turbulent at high Reynolds number, $C_D$ is of order unity. 
Fujiyama {\it et al.} estimated the drag force for the cases such as those in Fig. \ref{Fujiyama}.
The amplified line length can be related to the increase in energy, which should be equivalent to the work by the sphere. 
The drag coefficient $C_D$ obtained by these considerations was of order unity.
Thus Fujiyama {\it et al.}  could confirm an analogy between CT and QT for this situation too.

These numerical simulations seem to grab some essence of the phenomena, but not yet satisfactory.
For example, one of the important issues is lifetime of turbulence, which is a serious problem in CT \cite{Hof08}.
The lifetime is also observed in QT too \cite{Schoepe04,Yano10}.
Schoepe observed switching intermittency between turbulent and laminar phases around an oscillating microsphere, thus observing the lifetime of turbulence of the order of 100s \cite{Schoepe04}. 
Yano {\it et al.} could prepare two types of vibrating wires \cite{Yano10}.
One wire with remnant vortices works as a generator to make a vortex tangle around it.
The tangle continues to emit vortex loops, which is received by the other wire free from remnant vortices.
Even if we stop the vibration of the "generator" wire, the "receiver" continues to receive the loops for a period.
By measuring the average of the period, we can know the mean lifetime of QT created by the generator. 
The mean lifetime increases exponentially with the injection power of the generator to reach the order of 10$^3$s.  
However, the above numerical simulation \cite{Goto2008,Fujiyama2009} cannot reproduce this behavior, especially such a long lifetime.
If we stop supplying vortices to the oscillating sphere, the tangle around it disappears immediately. 
Currently we do not know any reason of the failure. 
One reason may be the surface roughness.
Usual numerical simulation supposes that the surface of solid is smooth for simplicity.
If the surface is rough, some pinning cites may trap or disturb the vortices and thus extend the lifetime.
However, the failure may be attributable to not such a boundary effect but some more serious bulk effect.

\subsubsection{Visualization of quantized vortices and turbulence}
 There has been little direct experimental information regarding the flow in superfluid $^4$He. This is mainly because usual flow visualization techniques are not applicable to cryogenic superfluids. However, this situation is rapidly changing \cite{SciverPLTP}. For QT, one can seed the fluid with tracer particles in order to visualize the flow field and possibly quantized vortices, which are observable by appropriate optical techniques. Polymer particles, solid hydrogen particles and metastable helium molecules are currently available as tracer particles.
 
 The first significant contribution was made by Zhang and Van Sciver \cite{Zhang05a}. Using a particle image velocimetry (PIV) technique with 1.7-$\mu$m-diameter polymer particles, they visualized a large-scale turbulent flow both in front of and behind a cylinder in a counterflow in superfluid $^4$He at finite temperatures. 
The second significant contribution was the visualization of quantized vortices by Bewley {\it et al.} \cite{Bewley06}.  In their experiments, the liquid helium was seeded with solid hydrogen particles smaller than 2.7 $\mu$m at a temperature slightly above $T_\lambda$, after which the fluid was cooled to below $T_\lambda$.  When the temperature was above $T_\lambda$, the particles were seen to form a homogeneous cloud that dispersed throughout the fluid. However, on passing through $T_\lambda$, the particles coalesced into web-like structures. Bewley {\it et al.} suggested that these structures represent decorated quantized vortex lines. 
Using the same technique, Paoletti {\it et al.} obtained the trajectories of tracer particles to visualize thermal counterflow \cite{Paoletti08}. The observed trajectories showed two distinct types of behavior. One group consisted of trajectories that moved in the direction of the heat flux, while the other consisted of those that opposed this motion. The former trajectories were smooth and uniform, but the latter could be quite erratic. Particles of the former trajectories were probably dragged by the normal fluid, while those of the latter were trapped in vortex tangles.
Recently metastable $^4$He$_2^*$ triplet molecules, with a radioactive lifetime of about 13 s in liquid helium, are available by a laser-induced-fluorescence technique. These molecules can act as tracers that follow the normal fluid flow \cite{Guo10}.
Their small size ($\sim$ 1 nm) means that they are not trapped by quantized vortices at temperatures above 1K. 
Guo {\it et al.} confirmed that the normal fluid flow is turbulent  too at relatively high velocities.

There are several theoretical and numerical efforts for revealing the complicated phenomena.
One of the key issues is to understand whether such tracer particles follow the normal flow or the superflow, or an even more complex flow. Poole {\it et al.} studied this problem theoretically and numerically, and showed that the situation changes depending on the size and mass of the tracer particles \cite{Poole05}. However, the situation is so complicated that many issues remain unanswered \cite{Sergeev06,Kivotides07,Kivotides08,Barenghi09}.
Mineda {\it et al.} formulated the coupled dynamics of fine particles and quantized vortices in the presence of a relative motion of the normal and superfluid components \cite{Mineda13}. 
Their numerical simulations agree with the velocity distributions of the tracer particles observed in thermal counterflow \cite{Paoletti08}.

\subsection{Recent developments in atomic BECs}
The achievement of Bose--Einstein condensation in trapped atomic gases in 1995 has stimulated intense experimental and theoretical activity in modern physics \cite{Pethick2008,Pitaevskii2003}.
This innovation has provided another important stage of quantum hydrodynamics because of the peculiar features of this system.
First, this system is a diulte Bose gas, thus the GP equation (\ref{gpeq}) gives a quantitatively accurate description of the static and dynamic properties of the atomic Bose-Einstein condensate(BEC). 
The vortex structure and dynamics can be discussed by a more fundamental approach than with superfluid helium. 
Secondly, the finite size effect due to the trapping potential and the associated density inhomogeneity 
yield new characteristics of vortex physics. 
Thirdly, the manipulation of a condensate wave function via external fields provides a versatile scheme to control the BEC. 
Fourthly, the vortex cores can be directly visualized through the observed density profile by a time-of-flight (TOF) technique. 
The TOF technique involves switching off the trapping 
field and taking an image of the BEC several milliseconds later. Switching off the trap allows the atomic gas to expand until a laser beam probe becomes available to observe the density profile. 
Fifthly, it is possible to load and cool atoms in more than one hyperfine spin state or more than one atomic element in the same trap, thus multicomponent BECs can be created experimentally. 
Multicomponent condensates allow the formation of various unconventional topological defects 
with complex properties that arise from interactions between different components \cite{Kasamatsurev}. 
This content relates closely to other fields of physics, such as superfluids $^{3}$He \cite{Vollhardt}. 
Finally, the most salient feature of the cold atom system is that a field-induced Feshbach resonance can tune 
the s-wave scattering length between atoms, which determines the strength of atom-atom interaction.
A Feshbach resonance occurs when a quasi-bound molecular state in a closed channel has an energy 
equal to that of two colliding atoms in an open channel. This technique can change the 
scattering length over a wide range, from negative values to positive ones.

Observing quantized vortices has been certainly an important smoking gun of superfluidity in atomic BECs.
Many experimental and theoretical works have been devoted to the issues of quantized vortices in this system \cite{Fetterrev2,KasamatsuPLTP, FetterJLTP,AndersonJLTP}.
Generally, we have two kinds of cooperative phenomena of quantized vortices, namely a vortex array under rotation and a vortex tangle in QT.
Both phenomena have been intensively studied in the field of superfluid helium, while most studies of quantized vortices in atomic BECs have been limited to the case of a small number of vortices or a vortex array under rotation.
Here we will briefly review the recent studies of QT in atomic BECs.
We will discuss three-dimensonal QT, followed by the review of QT in multi-component BECs.
Two-dimensional QT is currently another hot topic in this field \cite{Numasato10,Bradley12,Reeves12,Reeves13}, which cannot be addressed in this article. 

\subsubsection{Three-dimensional quantum turbulence}
In superfluid helium a dc velocity flow is usually applied to cause hydrodynamic instability and turbulence, which is not available in trapped atomic BECs.
Several methods of how to produce QT are proposed.  

\begin{figure}[ht]
\includegraphics[height=0.3\textheight]{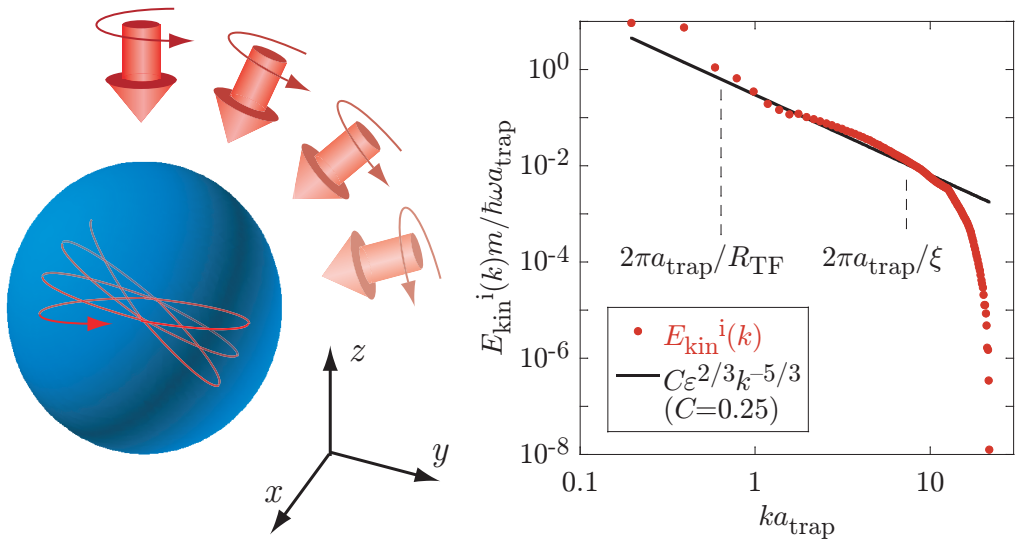}
\caption{QT in atomic BECs. 
(a) A method for realizing steady turbulence in an atomic BEC subject to precession. 
A BEC is trapped in a weakly elliptical harmonic potential. 
A rotation is applied along the $z$-axis followed by a rotation along the $x$-axis. 
(b) Energy spectrum of steady QT obtained by two-axis rotation. 
The dots represent the numerically obtained spectrum for an incompressible kinetic energy, 
while the solid line is the Kolmogorov spectrum. Here, $R_{\rm TF}$ is the 
Thomas--Fermi radius and $a_{\rm trap}=\sqrt{\hbar/m\omega}$ is the characteristic length scale of the trapping potential with the frequency $\omega$ [Kobayashi and Tsubota: Phys. Rev. A {\bf 76}, 045603 (2007), reproduced with permission.
Copyright 2007 by the American Physical Society].}
\label{twoaxispros}
\end{figure}

Berloff and Svistunov suggested the realization of QT in the dynamics of the formation of a BEC from a strongly degenerate nonequilibrium gas in which the phase of the macroscopic wave function was distributed randomly \cite{Berloff2002}.
Parker and Adams suggest the emergence and decay of turbulence in an atomic BEC under a simple rotation, starting from a vortex-free equilibrium BEC \cite{Parker}.
Starting from the vortex-free steady state with a constant potential, they performed a numerical simulation of the GP model with realistic experimental parameters for $^{87}$Rb BECs; the BEC performs a weak elliptical deformation to lead to turbulence with a Kolmogorov energy spectrum of Eq. (\ref{K41}).
Kobayashi and Tsubota proposed combining rotations around two axes, as shown in Fig. \ref{twoaxispros}(a) \cite{Kobayashi2007}.
Starting from a vortex-free initial state, they performed numerical simulations of the GP model with realistic parameters for $^{87}$Rb BECs, and obtained a steady turbulent state with no crystallization, but with highly-tangled quantized vortices.
The incompressible kinetic energy spectrum satisfies the Kolmogorov law in the inertial range (Fig. \ref{twoaxispros}(b)).

Currently there is only one experimental work successfully creating and observing three-dimensional turbulence. 
Henn {\it et al.} introduced an external oscillatory potential to an $^{87}$Rb BEC \cite{Henn2009a,Henn2009b}.
This oscillatory field induced a successive coherent mode excitation in a BEC.
They observed that increasing the amplitude of the oscillating field and the excitation period increased the number of vortices and eventually lead to the turbulent state \cite{Henn2009b}.
Henn {\it et al.} discovered another phenomenon probably characteristic of turbulence.
A usual BEC cloud is known to invert its own aspect ratio of the shape when it is allowed to expand freely because of the uncertainty principle \cite{Pethick2008}.  
However, a turbulent cloud was observed to expand keeping the aspect ratio \cite{Henn2009b}.
This behavior should come from the vortex configuration in a turbulent cloud \cite{Caracanhas13,Tsuchitani13}, but the mechanism is not yet revealed fully.

\subsubsection{Quantum turbulence in multi-component BECs}
Multicomponent BECs yield a rich variety of superfluid dynamics. We cannot cover all those topics in this article; the readers should refer to other review articles \cite{Novelsuperfluids,PR}.
This subsection reviews briefly binary QT in two-coponent BECs and spin turbulence in spinor BECs.

The macroscopic wave function $\Psi_j(\mathbf{x},t) (j=1,2)$ of two-component BECs at zero temperature obey the coupled GP equation \cite{Pethick2008}  
\begin{equation}
i \hbar \frac{\partial \,\Psi_j(\mathbf{x},t)}{\partial \,t} = \biggl( - \frac{\hbar ^2}{2m_j}\nabla^2
+\sum_k g_{jk} |\Psi_k(\mathbf{x},t)|^{2}-\mu_j \biggr) \Psi_j(\mathbf{x},t),
\label{2gpeq}
\end{equation}
where $m_j$ and $\mu_j$ are the particle mass and the chemical potential of the $j$th component. 
When the intracomponent interaction $g_{jj}$ and the intercomponent interaction $g_{jk}$ satisfy the relation $g_{11}g_{22}>g_{12}^2$, the two BECs like to be miscible.

When two-component BECs coexist with a relative velocity, they 
exhibit dynamic instability above a critical relative velocity \cite{Law}. 
This phenomenon is known as a {\it counter-superflow instability} (CSI). 
The CSI can be understood from the Bogoliubov spectrum for a uniform two-component BEC with a relative velocity \cite{Law}.
Takeuchi {\it et al.} suggested that the nonlinear dynamics triggered by the CSI  generates a binary QT \cite{TakeuchiCSI,Ishino}. 
\begin{figure}[ht] \centering
\includegraphics[height=0.11\textheight]{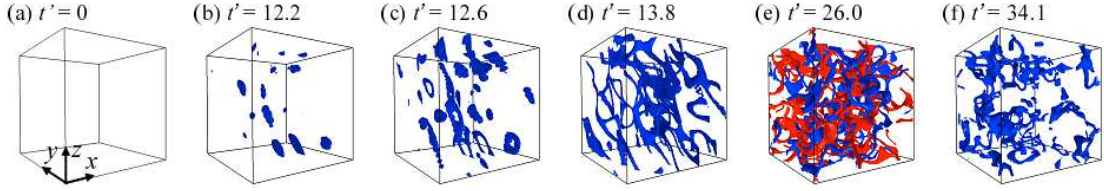}
\caption{Characteristic nonlinear dynamics of CSI of two-component BECs with a normalized time $t'$.
The parameters are set to $m_1=m_2=m$, $g_{11}=g_{22}=g$, and $g_{12}=0.9g$, similar to those in experiments. 
In the numerical simulation, the initial state is prepared by adding small random noise to the stationary wave function $\Psi_{i0}$ with $n_{10}=n_{20}=n_{0}$ and ${\mathbf V}_1=-{\mathbf V}_2$.
The panels shows the time development of low-density isosurfaces with $n_1({\mathbf x})=0.05n_0$ for ${\mathbf V}_1=\frac{1}{2}V_{R}\hat{\mathbf x}$.
Because of the symmetric parameter setting, the second component behaves in a similar manner to the first component (not shown).
Binary quantum turbulence is shown in (e). 
[Takeuchi {\it et al.}: Phys. Rev. Lett. {\bf 105}, 205301 (2010), reproduced with permission.
Copyright 2010 by the American Physical Society].}
\label{CSIfig}
\end{figure}
Figure \ref{CSIfig} depicts the characteristic nonlinear dynamics 
of the CSI in a uniform two-component BEC obtained 
by numerically solving the GP equation (\ref{2gpeq}).
In the early stage of the dynamics,
amplification of the unstable modes creates disk-shaped low-density regions
that are orientated in the flow direction [Fig. \ref{CSIfig}(b)].
The lowest density inside the disk region reaches zero, creating a local dark soliton.
The soliton in the $j$th component transforms into a vortex ring via snake instability
with a momentum antiparallel to the initial velocity [Fig. \ref{CSIfig}(c)].
These vortex rings grow to develop to turbulence.
The same scenario proceeds in both components, eventually leading to binary quantum turbulence [Fig. \ref{CSIfig}(e)].

The interesting point of this phenomenon would be how two BECs interact. Each BEC exhibits superfluidity respectively.
After vortex rings are nucleated and amplified,  they start to "communicate" by exchanging momentum via vortices, and thus the counterflow decays.
This situation is similar to thermal counterflow \cite{Adachi10} which has long been studied in superfluid $^4$He. 
There is a serious difference between thermal counterflow and CSI.
In thermal counterflow inviscid superfluid component and viscous normal fluid component interact via mutual friction through quantized vortices, thus leading to QT of superfluid component.
CSI lets two superfluids interact to develop to QT in both components.
It would be interesting to study how the energy spectra of each component is affected by their interaction.
The CSI is realizable experimentally \cite{Hamner}.

An optical trap removes the restriction of the confinable hyperfine spin states 
of atoms, thus realizing new multicomponent BECs with internal degrees 
of freedom, referred to as ``spinor BECs" \cite{Kawaguchireview,Kurnreview}. 
In contrast to the two-component BECs, interatomic interactions allow 
for a coherent transfer of population 
between different hyperfine spin states (spin-exchange collisions), which yields a fascinating physics different from two-component BECs. 

We consider a spin-1 spinor BEC at zero temperature.
 The macroscopic wave functions $\Psi _m$ with the magnetic quantum number $m$ ($m = 1,0,-1$) obey the GP equation \cite{Ohmi98, Ho98}
 \begin{eqnarray}
 i\hbar \frac{\partial}{\partial t} \Psi _{m} &=&  (-\frac{\hbar ^2 }{2M} \nabla ^2 + V) \Psi _{m} 
 +  \sum _{n=-1} ^{1} [- g \mu _{B}(\mathbf{B} \cdot \hat{\mathbf{S}})_{mn} + q(\mathbf{B} \cdot \hat{\mathbf{S}})^{2}_{mn} ]\Psi _{n}  \nonumber \\
 &+& c_{0} n \Psi _{m} +  c_{1} \sum _{n=-1} ^{1} \mathbf{s} \cdot \hat{\mathbf{S}} _{mn} \Psi _{n}.  \label{2dGP}
 \end{eqnarray}
Here, $V$ and $\mathbf{B}$ are the trapping potential and magnetic field. The parameters $M$, $g$, $\mu _{B}$, and $q$ are the mass of a particle,
the Land$\rm \acute{e}$ $g$ factor, the Bohr magneton, and a coefficient of the quadratic Zeeman effect, respectively.
 The total density $n$ and the spin density vector $s_{i}$ ($i = x, y, z$ ) are given by $n =  \sum _{m=-1} ^{1}|\Psi _m|^2$ and  $s_{i} = \sum _{m,n = -1}^{1} \Psi _{m}^{*} (\hat{S}_{i})_{mn} \Psi _{n}$ with
 the spin-1 matrices $(\hat{S}_{i})_{mn}$. The parameters $c_{0}$ and $c_{1}$ are the coefficients of the spin-independent and spin-dependent interactions.
We focus on the spin-dependent interaction energy $E_{spin} = \frac{c_{1}}{2} \int \mathbf{s} ^{2} d \mathbf{r}$, whose coefficient $c_{1}$ determines whether the system is ferromagnetic ($c_{1} < 0$) or antiferromagnetic ($c_{1} > 0$).
For simplicity we describe the ferromagnetic case.
The antiferromagnetic case gives another, just more complicated story \cite{FT13}.

\begin{figure}[ht] \centering
\includegraphics[height=0.17\textheight]{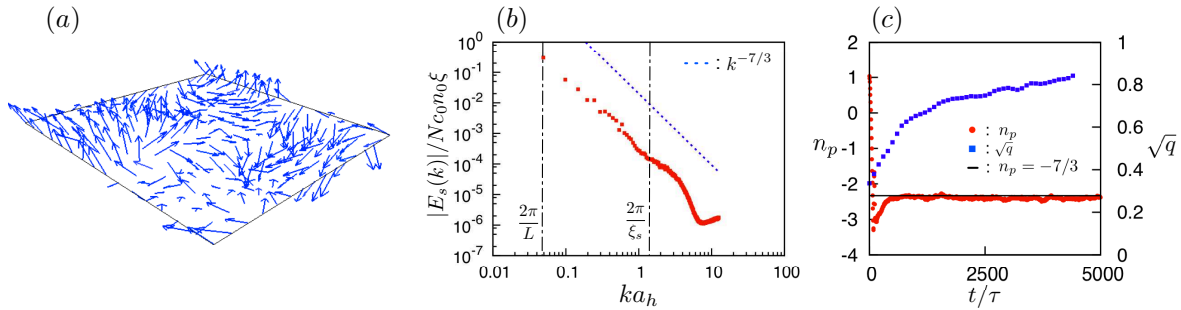}
\caption{Spin turbulence in spinor BECs. The dynamics starts from the initial state of the counterflow between the $m=\pm 1$ components. 
(a) Distribution of spin density vectors $ \mathbf{s} $ in ST at $t/\tau = 4000$. (b) Spectrum of the spin-dependent interaction energy  $t/\tau = 4000$ with a characteristic time $\tau$. (c) Time development of the exponent $n_{p}$ of the spectrum (b) and the spin glass order parameter $ \sqrt{q(t)} $.  See  Ref. \cite{TAF13} for the detail.}
\label{ST}
\end{figure}

When the system is highly excited from the ground state, it goes through hydrodynamic instability to spin turbulence (ST) in which the spin density vector has various direction.  
Figure \ref{ST} shows the behavior of ST starting from the initial state of the counterflow between the $m=\pm 1$ components in a uniform two-dimensional system \cite{FT12a,TAF13}.
When the spin density vectors are randomly distributed (Fig. \ref{ST} (a)), the spectrum of the spin-dependent interaction energy becomes to obey a $-7/3$ power law  (Fig. \ref{ST} (b)); 
the $-7/3$ power law is understood by the scaling analysis of the time-development equation of the spin density vector \cite{FT12a}. 
This $-7/3$ power law is so robust that it is confirmed numerically also in a trapped system with the instability of the initial helical structure of spins \cite{FT12b} and a uniform system excited by an oscillating magnetic field \cite{AT13}.  
Another characteristic of ST is that the spin density vectors become spatially random but temporally frozen, which reminds us of spin glass.
Spin glasses are magnetic systems in which the interactions between the magnetic moments are in conflict with each other \cite{RMP}.
Thus, these systems have no long-range order but exhibit a freezing transition to a state with a kind of order in which the spins are aligned in random directions.

In order to characterize the spin-glass-like behavior in ST, Tsubota {\it et al.} introduced an order parameter following the spin-glass theory \cite{RMP,SK}.
We should note that the order parameter is time-dependent.
For the normalized spin density vector  $\hat{\mathbf{s}}(\mathbf{r},t)=\mathbf{s}(\mathbf{r},t)/|\mathbf{s}(\mathbf{r},t)|$, its space average and the time average during the period $[t, t+T]$ could be defined as 
\begin{equation}
 [\hat{\mathbf{s}} (\mathbf{r},t)]=\frac1A \int_A \hat{\mathbf{s}} (\mathbf{r},t)d\mathbf{r}, \quad  \langle \hat{\mathbf{s}} (\mathbf{r},t)\rangle_T= \frac1T \int_t^{t+T}\hat{ \mathbf{s}} (\mathbf{r},t_1)dt_1
 \label{average}\end{equation}
with  the system area $A$. The argument of how to take $T$ is discussed in Ref. \cite{TAF13}.
The time-dependent order parameter is defined as 
\begin{equation}
q(t)=[\langle\hat{ \mathbf{s}} (\mathbf{r},t)\rangle_T^2]. \label{q}
\end{equation}
 Figure \ref{ST} (c) shows the time dependence of $q(t)$ and the power exponent $n_p(t)$ of the spectrum of the spin-dependent interaction energy. 
The order parameter $q(t)$ increases obviously as the exponent $n_{p}(t)$ approaches $-7/3$, which means that the spin-glass-like order grows as the ST with the $-7/3$ power law develops. 
Of course, our system of spinor BECs differs from a magnetic system yielding spin glass, 
and the spin turbulence is not spin glass. 
We do not know currently what causes the spin-glass-like behavior  \cite{TAF13}.

\section{Why is quantum turbulence interesting?} 
This section describes a few comments on why the research of QT is interesting, what the significance is.
They are challenging problems in this field. 

\subsection{Reductionism}
Comparing QT and CT reveals definite differences between them. 
Turbulence in a classical viscous fluid appears to be comprised of vortices, for example, as pointed out by da Vinci by his famous sketch. 
However, these vortices are unstable, repeatedly appearing and disappearing. 
Moreover, the circulation is not conserved and is not identical for each vortex. 
To the contrary, QT consists of a tangle of quantized vortices which are well-defined topological defects with the same conserved circulation.  
Looking back at the history of science, {\it reductionism}, which tries to understand the nature of complex things by reducing them to the interactions of their elementary parts, has played an extremely important role. 
The success of solid state physics owes much to  {\it reductionism}. 
In contrast, conventional fluid physics is not reducible to elements, and thus does not enjoy the benefits of {\it reductionism}. 
However, quantum turbulence is different, being reduced to quantized vortices. 
Thus {\it reductionism} is applicable to quantum turbulence. 
Consequently, QT should lead to a simpler model of turbulence than CT. 
For example, it is possible to connect directly the configuration of vortices and the spectrum of the kinetic energy; we may ask what kinds of configurations yield the Kolmogorov spectrum.
Such approaches are partly successful in two-dimensional quantum turbulence \cite{Bradley12, Kusumura13}. 
 
\subsection{Relation between the real space and the Fourie space}
In a statistically steady state of three-dimensional classical turbulence the energy spectrum yielding the Kolmogorov law is sustained by the self-similar energy cascade from large to small scales \cite{Frisch,Davidson}. 
And the energy cascade is believed to be connected with the Richardson cascade in which large eddies are successively broken up to smaller eddies. 
However, this picture is only phenomenological in CT, because eddies or vortices are not so well-defined.
The story is different in QT where the rotational velocity field is completely attributable to well-defined quantized vortices.
The energy cascade in the Fourie space should come from the cascade of vortices in the real space; it would be possible to confirm experimentally or numerically the "genuine" Richardson cascade in QT. 

\subsection{How to characterize QT; Introduction of the order parameters}
The definition of turbulence in a classical fluid is not so definite;  it may differ from a scientist to a scientist.
Then, how about QT?  Can we define and characterize QT by taking the above advantage?
QT consists of quantized vortices, but any random distribution of vortices dose not necessarily yield QT.
Focussing on some statistical laws or introducing an order parameter would be useful for the definition of QT.
The former effort is now in progress in the field of superfluid helium \cite{Walmsley08}.
The latter approach appears in introducing the order parameter of Eq. (\ref{q}) in spin turbulence.
If these approaches are successful, it contributes so much for understanding QT as well as CT.

Another current important approach is to consider QT as a meta-stable transient state, characterized  as a nonthermal fixed point.
Coming from the idea of critical phenomena, this approach expects that QT has universal properties such as self-similarity and independence of the details of how the respective state has been reached.
The wealth of universal phenomena in the far-from-equilibrium state could shed light on revealing QT \cite{Nowak12,Schole12,Karl13}.

\ack The author thanks Michikazu Kobayashi and Kazuya Fujimoto for preparing the manuscript.

\end{document}